# Software Based Pole-Zero Extraction Technique for $n^{th}$ Order Analog Filters

Arnesh Sen
*Dept. Of Physics*
*Jadavpur University*
Kolkata – India
Email: senarnesh.elec@gmail.com

Aishik Das
*Dept. of Electronics*
*University of Calcutta*
Kolkata - India
Email: aishik.das08@gmail.com

Prof. Jayoti Das
*Dept. Of Physics*
*Jadavpur University*
Kolkata - India
Email: jayoti.das@gmail.com

*Abstract*—The positions of poles and zeros of a filter circuit indicates how the system behaves with variation in frequency as well as it results a tool to utilize the system gain by setting the frequency range in a complex process. The existing solutions to extract the poles and zeros often include too complex calculations to use it frequently. In this paper we propose a compact tool to detect the position of poles and zeros of any practically designed analog filter of any order without going through the working principle of the circuit as well as by using this technique calculation of transfer function is doesn't need to do derive which often becomes hectic.

## I. INTRODUCTION

Poles and Zeros of a particular circuit allows us to examine how the transfer function varies with the variation of frequency. By examining the frequency response we can use that circuit according to our convenience. For instance various analog filters are designed and applied by judging their frequency response. The conventional way to extract the exact frequency response of a circuit is at first one have to calculate the transfer function and locations of poles and zeros of the circuit. Well, it sounds like just a piece of cake but in reality for higher order systems the situation becomes often pathetic when someone tries to do it manually. Now, the question may be arisen that then what are the uses of the simulators as they are made to make easy the manual effort. The answer is quite simple simulators can handle the complex physical systems in some extent but they have also limitations. Bode Plot [1] and Pole-Zero analysis method [2] are the name of the some of those existing method to calculate frequency response. One innovative and interesting approach to get instant rough idea of the frequency response of any particular circuit shown by Behzad Razavi is "Association of Poles with Nodes" [3]. It should be mentioned here all the existence process[4][5] needs prior understanding and solving skills should be developed to apply these process in an unknown circuit. But what about the calculation of transfer function. There is no general way to extract the transfer function from any arbitrary circuit. This paper propose a general software based method which is able to extract the transfer function using experimental data as well as to approximate the locations of poles and zeros of that circuit. Here MATLAB is the heart of the entire process. Some of the known filter circuit[6][7] is run with this process and the results are verified. Using this method one can get the approximate locations of poles and zeros without having any knowledge of the internal behavior of the circuit components and this key feature make this process unique and innovative.

## II. METHODOLOTY

Suppose we have an unknown circuit, it may be an amplifier or any arbitrary analog filter with any order. We just have to energies the circuit by providing any kind of input like step, ramp, impulse and sinusoid. It should be mentioned here that both the input and the output terminals of the circuit is interfaced with MATLAB software via microcontroller (Arduino is preferable). The output voltage response with variation of frequency is recorded and plotted by the analog interfacing facility which should be inbuilt in the microcontroller. The output voltage response can directly be interfaced with MATLAB but to detect input frequency one frequency to voltage converter with unit calibration factor is used.

Using the best-fit facility, available in MATLAB we can approximate the output voltage response. By dividing this response with input voltage we can get the response of the circuit gain and transfer function can easily be calculated. Here we don't need to worry about complex roots as MATLAB supports imaginary numbers. Now both the numerator and denominator are converted in factor form and from this kind of expression poles and zeros are extracted.

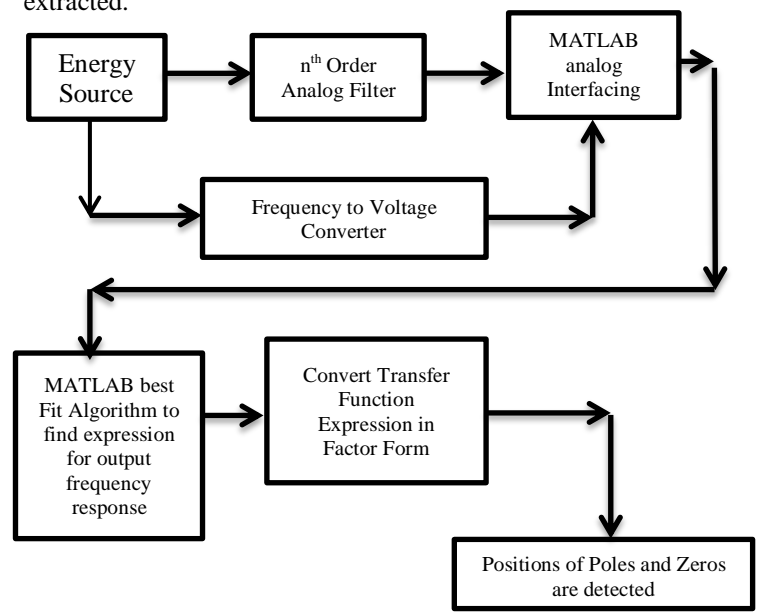

## III. DETAILS ANALYSIS OF FIRST ORDER HIGH PASS FILTER USING THIS METHOD.

According to this method we have to extract data from practical circuit (Figure-1). Real time simulation method is used here by applying analog read command for analog interfacing. After applying best fit method we get the polynomial expressions for output response. And by following the process described in the flow chart we have the transfer function we can easily locate the poles and zeros.

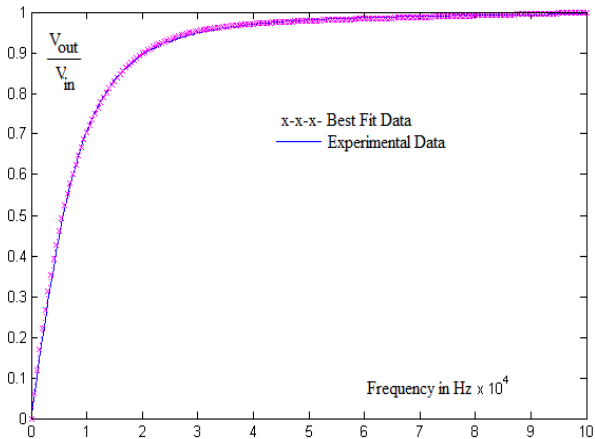

Figure-1: First order High-Pass Filter Frequency Response

The above figure is shown the comparison between the experimental data extracted directly from the experiment and MATLAB best fit output. Two plots are almost indistinguishable and that's why the equation, given by MATLAB best method can be allowed to move forward to next step.

For this example best fir equation are resulted as,

$y = a.\exp(bx) + c.\exp(dx)$ where, $a = 9616$;

$b = 3.833 \times 10^{-7}; c = -9616$ and $d = -0.0001303$

x and y represents angular frequency and gain respectively. Now after converting this expression in "numerator / denominator" form we get, (See Appendix)

$Transfer\ Function\ (T.F) = \frac{9652.54s}{s+10000}$ and expected pole-zero plot is shown in Figure - . This is as same as the pole-zero plot of high-pass filter.

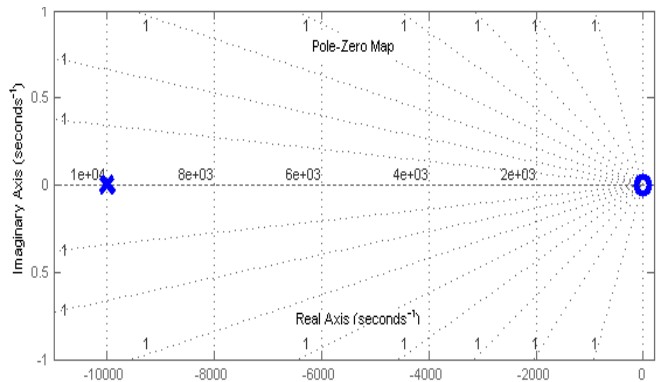

Figure-2: 1st Order High Pass Filter Pole-Zero Plot

It should be mentioned here that all the above calculations are shown for understanding purpose. While using this process the calculations will be done by MATLAB. But user must take the responsibility of the choice of best fit. For instance in the case of 1st order high pass filter the best match is found by using two-termed exponential function. The value of R and C is chosen as R=1k and C=0.1uF.

## IV. RESULTS AND DISCUSSIONS

We apply this method on arbitrary filter circuits to ensure the validity of the method. All the results are well matched with the conventional results. But choices of best fits are different.

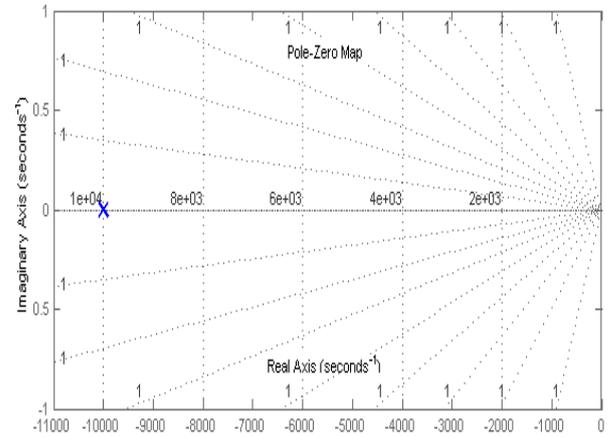

Figure-3: 1st Order Low Pass Filter Pole-Zero Plot

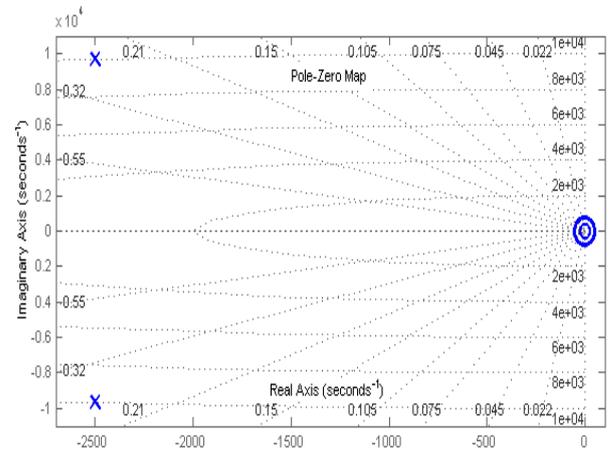

Figure-4: 2nd Order High Pass Filter Pole-Zero Plot

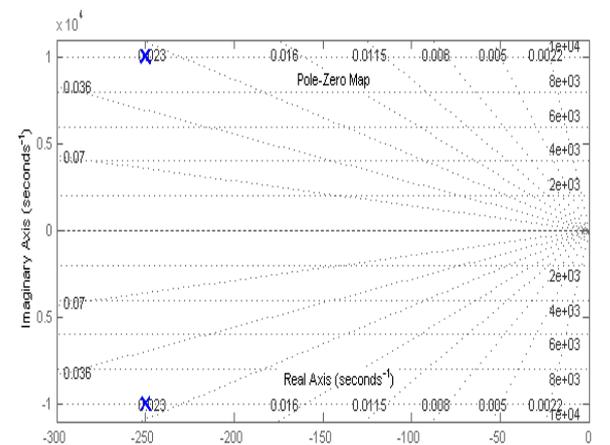

Figure-5: 2nd Order Low Pass Filter Pole-Zero Plot

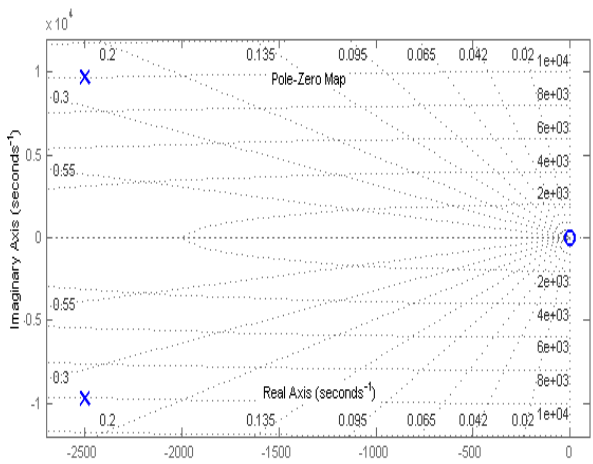

Figure-6: Band Pass Filter Pole-Zero Plot

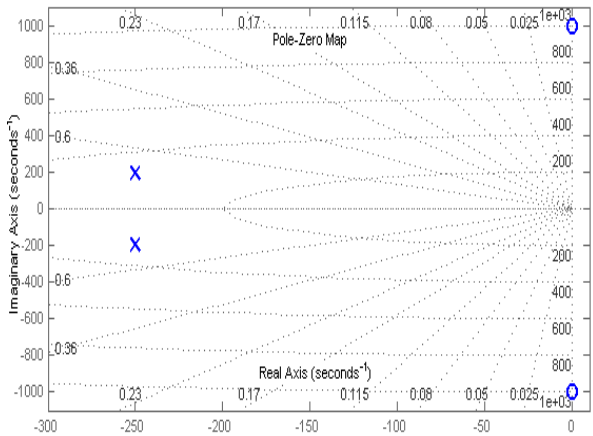

Figure-7: Notch Filter Pole-Zero Plot

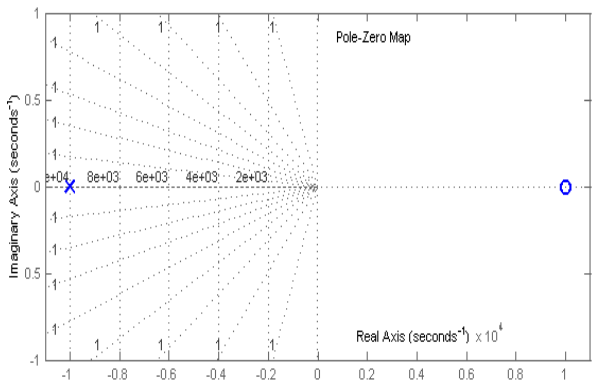

Figure-8 : 1st Order All – Pass Filter Pole-Zero Plot

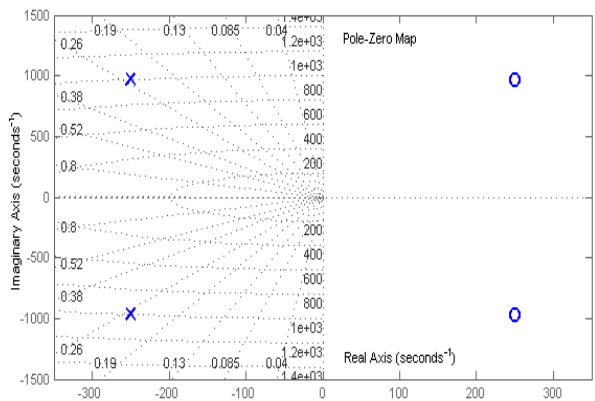

Figure-9 : 2nd Order All – Pass Filter Pole-Zero Plot

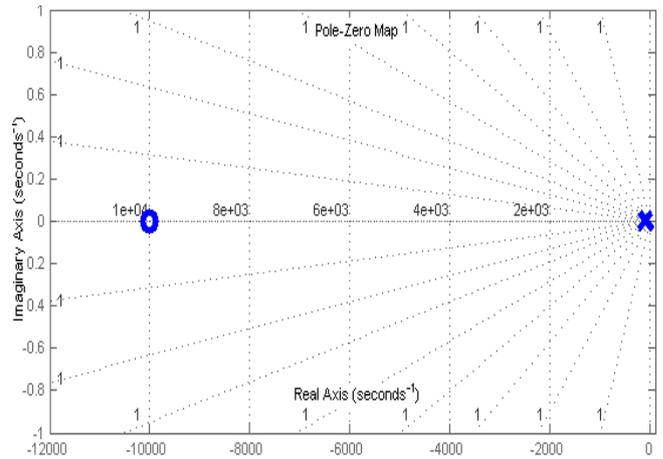

Figure-10: General Filter Pole-Zero Plot

The locations of poles and zeros located in the above diagram are approximately matched. Here the entire second order filter is best fitted in Gaussian Function. LM331 is used here as frequency to voltage converter [8].

## V. CONCLUSION

A novel approach is proposed to detect the locations of poles and zeros of any arbitrary analog filters of any order without understanding the working principle of the circuit. This process is entirely software based and the circuit must be practically designed and interfaced properly with MATLAB via microcontroller to utilize this tool.

## APPENDIX

$$y = 9616.\exp(3.8 \times 10^{-7}) - 9616.\exp(-0.0001x)$$

$$or, y = \frac{9616.\exp(3.8 \times 10^{-7}x + 0.0001x) - 9616}{\exp(0.0001x)}$$

$$or, y \approx \frac{0.9652x}{0.0001(x + 10000)}$$

$$or, y \approx \frac{9652.54x}{x + 10000}$$

ACKNOWLEDGMENT

We would like to show gratitude to Dr. Jayoti Das, Associate Professor, Dept. of Physics, Jadavpur University for her kind support, constant supervision and encouragement during this course of research work which will help us to reach intended goal in the midst of intermingled maze of possibilities and failures.